\documentclass[aps,pre,twocolumn,showpacs,a4paper,superscriptaddress]{revtex4-1}
\usepackage{graphicx,bm,amsmath,dcolumn,amssymb,color}
\graphicspath{{figures}}
\usepackage{geometry,hyperref}
\begin{document}

\title{Overlap of two topological phases in the antiferromagnetic Potts model}
  \author{Ran Zhao}
  \affiliation{Hefei National Laboratory for Physical Sciences at Microscale,
  Department of Modern Physics, University of Science and Technology of China, Hefei, 230027, China}
  \author{Chengxiang Ding}
  \email{dingcx@ahut.edu.cn}
  \affiliation{School of Science and Engineering of Mathematics and Physics, Anhui University of Technology, Maanshan 243002, China }
  \author{Youjin Deng}
  \affiliation{Hefei National Laboratory for Physical Sciences at Microscale,
  Department of Modern Physics, University of Science and Technology of China, Hefei, 230027, China}
  \date{\today}
\begin{abstract}
   By controlling the vortex core energy, the three-state ferromagnetic Potts model
can exhibit two types of topological paradigms, including the quasi-long-range ordered phase and the vortex lattice phase [PRL {\bf 116}, 097206 (2016)].
Here, by Monte Carlo simulations using an efficient worm algorithm, we show that by controlling the vortex core energy, the {\it antiferromagnetic} Potts model can also exhibit the two topological phases, 
more interestingly, the two topological phases can overlap with each other.
\end{abstract}
\maketitle 
%\end{CJK*}
\section{Introduction}
The famous Berezinskii-Kosterlitz-Thouless (BKT) transition, firstly found in the two dimensional $XY$ model\cite{XY1,XY2},
is a typical example of classical topological transition, which is not caused by the spontaneous breaking of symmetry. 
For all the temperatures below the BKT point, the system is critical, with algebraically decaying correlation function, 
and the critical exponent describing the behaviors of the correlation function varies with the temperature. 
Such phase is called quasi-long-range ordered (QLRO) phase,  which is a typical example of classical topological phase. 
The excitations in the model include spin waves and vortices, 
among which the latter plays a key role in the thermodynamic properties of the system.
In fact, some models with discrete symmetry of spins, can also have such topological excitations, 
and consequently the QLRO phase and BKT transition, 
such as the clock model\cite{clock}, 
the finite-temperature quantum Ising antiferromagnet on triangular lattice\cite{QIsing},
or the classical antiferromagnetic Ising model on multilayer triangular lattice\cite{Ising}.
In addition, those models can also be ordered, and the transition between the ordered phase and the 
QLRO phase is also called BKT transition. % although it spontaneously breaks the symmetry.

Besides the aforementioned clock model and Ising model, the three-state antiferromagnetic Potts model (AFP) can also exhibit QLRO and 
BKT transition. The square-lattice three-state AFP\cite{afp3sq} have very similar critical properties
as the triangular-lattice antiferromagnetic Ising model, it is also critical at the zero temperature, 
although the critical exponent $\eta$ is different. On the multilayer lattice\cite{afp3layer,Ising}, 
the symmetry and critical properties of the two models are isomorphic, with emergent U(1) symmetry and QLRO at the intermediate temperature, sandwiched
by the low-temperature ordered phase breaking $Z_6$ symmetry and the high-temperature disordered phase. Similar properties can also be found in
the square-lattice AFP with next nearest neighboring ferromagnetic interactions\cite{Nijs}.
%By the theoretical analysis\cite{Nijs}, the critical properties of the AFP model are really controlled by the vortex exitations in the model,
% which can also be detected by numerical simulations
In Ref. \onlinecite{Nijs}, the vortex excitations and their decisive effect on thermodynamic properties are identified by theoretical analysis;
In Refs. \onlinecite{afpvortex1,afpvortex2}, the vortex excitations are directly investigated by numerical simulations.

\begin{figure}[htpb]
\includegraphics[width=\columnwidth]{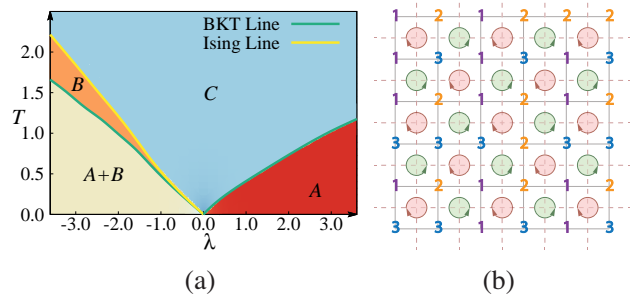}
\caption{(Color online) (a) Phase diagram of the generalized antiferromagnetic Potts model (\ref{Potts}),
 where A is the quasi-long-range ordered phase, B is the vortex lattice phase, and C is the disordered phase.
 (b) A snapshot of the phase A+B; the cycles, added to guides eyes, denote the vortices on the dual lattice. }
\label{PD}
\end{figure}

Recently, by Monte Carlo simulations, Bhattacharya and Ray\cite{Ray} show that the QLRO phase can also be found
in the ferromagnetic three-state Potts model if the vortex excitations are suppressed by controlling the core energy of the vortices.
In addition, another classical topological phase, the vortex lattice, can also be found in the model if the vortex excitations are enhanced. 
This profoundly reveals the nature of the model, indicating the importance of the topological excitations.
In current paper, we study the antiferromagnetic Potts model by the similar way, with an efficient worm algorithm. 
We find that the QLRO phase can always be found in the phase diagram (Fig. \ref{PD}) 
no matter the vortex excitation is suppressed or enhanced.
Furthermore, we find that the QLRO phase can overlap with the vortex lattice phase; 
one is on the original lattice and the other is on the dual lattice.
As far as we know, the overlap of the two topological phases is reported for the first time.

\section{Model and Method}
We study the Potts model on the square lattice $\Lambda$
\begin{eqnarray}
\mathcal{H}=-J\sum\limits_{\langle i,j\rangle\in\Lambda}\delta_{\sigma_i,\sigma_j}
  +\lambda\sum\limits_{i^\prime\in\Lambda^\prime} |\omega_{i^\prime}|, \label{Potts}
 \end{eqnarray}
where $J$ is the coupling constant of the nearest-neighboring Potts spins $\sigma_i$ and $\sigma_j$,
which can take value 0, 1, or 2. 
The interaction is ferromagnetic one if $J=1$, and antiferromagnetic one if $J=-1$. In current paper, we focus on the 
antiferromagnetic case.
$\lambda$ is the energy of a vortex $\omega_{i^\prime}$, which is defined on the dual lattice $\Lambda^\prime$. 
$\omega_{i^\prime}=(\Delta_{ba}+\Delta_{cb}+\Delta_{dc}+\Delta_{ad})/3$, with $\Delta_{ba}=\sigma_b-\sigma_a$, where $\sigma_a$, $\sigma_b$,
$\sigma_c$, and $\sigma_d$ are the four Potts spins on the square plaquette in $\Lambda$, surrounding the site $i^\prime$ anticlockwise. 
The first term of the Hamiltonian accounts for the {\it pure} Potts model, $\lambda$ can control the vortex excitations of the model. For $\lambda>0$,
 the vortex excitation is suppressed; for $\lambda<0$, the vortex excitation is enhanced.
%for the antiferromagnetic case, the model is crtical at zero temperature, 
%demonstrating a classical `spin liquid' like the triangular lattice antiferromagnetic Ising model. Although the vortex exitations of the 
%pure AFP model has ever been detected by Monte Carlo simulations, QLRO phase is not found in the model. 

To simulate the model effectively by Monte Carlo method is not a trivial work, although the pure Potts model can be efficiently simulated 
by the Metropolis algorithm or the more efficient cluster algorithms\cite{SW,Wolff,WSK,Geo}, 
the general case of model (\ref{Potts}) has severe freezing problems
in Monte Carlo simulations. As pointed by Ref. \onlinecite{Ray}, the auto correlation is very large for the single spin-flip algorithm. 
To solve the freezing problem and also the problem of critical slowing down, we formulate a worm algorithm below.

\begin{figure}[htpb]
\includegraphics[width=0.7\columnwidth]{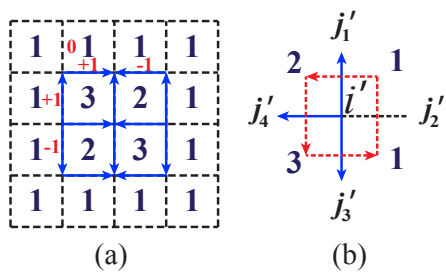}
\caption{(Color online) (a) Definition of the current; (b) Represent the vortex by current: 
$\omega_{i^\prime}=(f_{i^\prime,j^\prime_1}+f_{i^\prime,j^\prime_2}-f_{i^\prime,j^\prime_3}-f_{i^\prime,j^\prime_4})/3$.}
\label{config}
\end{figure}

For each pair of nearest neighboring sites $i^\prime$ and $j^\prime$ on the dual lattice, 
we define current $f_{i^\prime,j^\prime}$, whose value is determined by the corresponding 
spins $\langle \sigma_{i}, \sigma_{j}\rangle$ on the original lattice.
When $\sigma_i\ne\sigma_j$, the value of $f_{i^\prime,j^\prime}$ is +1 if the direction of the
arrow linking sites $i^\prime$ and $j^\prime$ is upward or rightward, conversely, it is -1.
The direction of the arrow is determined by the following rule: the left spin $\sigma_i$ of the arrow should be higher than 
the right spin $\sigma_j$, with $1<2<3<1$. 
When $\sigma_i=\sigma_j$, the value of $f_{i^\prime,j^\prime}$ is zero, and the linking between $i^\prime$ and 
$j^\prime$ is a straight line. Some concrete examples are shown in Fig. \ref{config} (a).
The vortex can also be written as the combination of the current $\omega_{i^\prime}=f_{i^\prime}/3$, 
with $f_{i^\prime}=f_{i^\prime,j^\prime_1}+f_{i^\prime,j^\prime_2}-f_{i^\prime,j^\prime_3}-f_{i^\prime,j^\prime_4}$ the total current 
flowing out of site $i^\prime$, as shown in Fig. \ref{config} (b).
Because the total current $f_{i^\prime}$ is always zero modular 3, i.e., $\mod(f_{i^\prime},3)=0$, 
the value of $\omega_{i^\prime}$ can only be an integer 0 or $\pm 1$. %that are identified as vortex or anti-vortex, respectively.

The $\delta$ function can be rewritten as $\delta_{\sigma_i,\sigma_j}=1-|f_{i^\prime,j^\prime}|$,
thus the Hamiltonian can be written as
\begin{eqnarray}
\mathcal{H}^\prime=-2NJ+J\sum\limits_{\langle i^\prime,j^\prime\rangle\in\Lambda^\prime}|f_{i^\prime,j^\prime}|
  +\sum\limits_{i^\prime\in\Lambda^\prime}|\omega_{i^\prime}|. \label{current}
\end{eqnarray}
Our worm algorithm is based on $\mathcal{H}^\prime$. It should be noted that the correspondence between 
the current model and the spin model is not one-by-one, 
there exists some current configurations those have no correspondence in spin configurations, due to the periodic boundary conditions.
 The computation of physical quantities can only be done in the current configurations which have correspondence in spin configurations.
The worm algorithm is described below
\begin{enumerate}
\item Randomly choose Ira=Masha in the sites of the dual lattice, randomly choose an updating scheme $\otimes=+$ or $-$;
\item\label{b} Exchange Ira and Masha and set Ira as the moving head of the worm;
\item Randomly choose one of the nearest neighboring sites of Ira as $j^\prime$;
\item Propose to update the current of bond $\langle {\rm Ira}, j^\prime\rangle$ 
      by $f^\prime_{{\rm Ira},j^\prime}=f_{{\rm Ira},j^\prime}\otimes 1$, 
      using rules: -1+1=0, 0+1=1, 1+1=-1; -1-1=+1, 0-1=-1, +1-1=0;
\item Accept the proposed change with probability $p=\min(1,R=e^{-\Delta H/T})$;
\item If Ira=Masha, exit; otherwise go back to (\ref{b}).
\end{enumerate}

\begin{table}[htbp]
\caption{Auto correlation time of the Metropolis algorithm and the worm algorithm for the generalized Potts model (\ref{Potts})
 at a BKT point ($T=1.06$, $\lambda=-2$).}
 \begin{tabular}{c|c|c|c|c|c}
    \hline
                 $L$   &~16~~&~32~~   & ~64~~  &~128~   & ~256~\\
    \hline
 $\tau$ of Metropolis   &29.5 & 97.4  & 443.2  &1647.3  &6200.1 \\
    \hline
 $\tau$ of worm        &24.8 &30.8    & 39.2   & 48.4   & 62.7 \\
    \hline
\end{tabular}
\label{tau}
\end{table}
We have compared the efficiency of the worm algorithm to the Metropolis algorithm. 
Firstly, we test at the low temperature (with $T<=0.5$,  $\lambda=-2$, and $L>=64$), we find that for different random seeds, 
the Metropolis algorithm may give obviously different values of given physical quantity, 
although each average is over millions of samples and million sweeps are abandoned for thermalization. 
This means the Metropolis algorithm may be trapped in a local minima of the free energy. 
The above symptom of freezing problem does not appear in the simulations of the worm algorithm;
 furthermore, the sweeps for thermalization is much less than the Metropolis algorithm, generally thousands of sweeps is enough.
 Secondly, at relatively high temperature, Metropolis also works, but the auto correlation time is much larger than the worm algorithm; 
take the critical point of a BKT transition as example ($T=1.06$ and $\lambda=-2$), for the large system, 
the autocorrelation time of Metropolis is hundreds times of worm, as shown in Table \ref{tau}.
From the data, we can fit the dynamical exponent of the two algorithms by the formula:
\begin{eqnarray}
\tau=\tau_0+aL^z,
\end{eqnarray}
with $\tau$ the correlation time and $z$ the dynamical exponent.
The fitting gives $z_{\rm Metropolis}=1.9(2)$ and $z_{\rm worm}=0.3(1)$, this means the critical slowing down is sharply reduced by the worm algorithm.

The sampled quantities include the density of domain wall $\rho_{dm}$, density of vortices $\rho_{vx}$, the specific heat $C_{dm}$ and $C_{vx}$ corresponding to the two types of energy, respectively
\begin{eqnarray}
\rho_{dm}&=&\frac{1}{L^2}\sum\limits_{\langle i,j\rangle\in \Lambda} (1-\delta_{\sigma_i,\sigma_j}),\\
\rho_{vx}&=&\frac{1}{L^2}\sum\limits_{i^\prime\in\Lambda^\prime} |\omega_{i^\prime}|,\\
C_{dm}&=&\frac{L^2(\langle\rho_{dm}^2\rangle-\langle\rho_{dm}\rangle^2)}{T^2},\\
C_{vx}&=&\frac{L^2(\langle\rho_{vx}^2\rangle-\langle\rho_{vx}\rangle^2)}{T^2},
\end{eqnarray}
where $T$ is the temperature.

We also sample the staggered magnetization $m_s$, the staggered vortex magnetization $m_{vx}$, and the Binder ratios $Q_s$
and $Q_{vx}$ corresponding to the two types of magnetizations, respectively
\begin{eqnarray}
m_{s}&=&\langle|\mathcal{M}_s|\rangle,\\
m_{vx}&=&\langle|\mathcal{M}_{vx}|\rangle,\\
Q_s&=&\frac{\langle \mathcal{M}_s^2\rangle^2}{\langle \mathcal{M}_s^4\rangle},\\
Q_{vx}&=&\frac{\langle \mathcal{M}_{vx}^2\rangle^2}{\langle\mathcal{M}_{vx}^4\rangle},
\end{eqnarray}
where
\begin{eqnarray}
\mathcal{M}_s&=&\frac{1}{L^2}\sum\limits_{\vec{r}\in \Lambda} (-1)^{x+y}\vec{s}(\vec{r}), \\
\mathcal{M}_{vx}&=&\frac{1}{L^2}\sum\limits_{\vec{r}^\prime\in\Lambda^\prime}(-1)^{x^\prime+y^\prime}\omega_{\vec{r}^\prime},
\end{eqnarray}
with $\vec{s}(\vec{r})$ the vector mapped from the Potts spins by the rule
$\vec{s}(\vec{r})=\exp\{2\pi i\sigma(\vec{r})/3\}$. $(x,y)$ and $(x^\prime,y^\prime)$
are the coordinations of sites $\vec{r}$ and $\vec{r}^\prime$, respectively.
$m_s$ can detect the staggered order on the original lattice, while the $m_{vx}$ can detect the vortex lattice order on the dual lattice.

Another important quantity that we sampled is the correlation length $\xi$
\begin{eqnarray}
\xi=\frac{(\chi/F-1)^{1/2}}{2\sqrt{\sum\limits_{i=1}^d\sin^2(\frac{k_i}{2})}},
\end{eqnarray}
where $\vec{k}$ is the ``smallest wave vector" of the square lattice along the 
$x$ direction, i.e., $\vec{k}\equiv (2\pi/L,0)$. The susceptibility $\chi$ and the 
``structure factor" $F$ are 
\begin{eqnarray}
\chi&=&\frac{1}{L^2}\langle\big|\sum\limits_{\vec{r}}(-1)^{x+y}\vec{s}(\vec{r})\big|^2\rangle,\\
 F&=&\frac{1}{L^2}\langle|\sum\limits_{\vec{r}} (-1)^{x+y}e^{i\vec{k}\cdot\vec{r}} \vec{s}(\vec{r})|^2\rangle.
\end{eqnarray}
$\xi$ is an important quantity that can identify the type of phase. In the disordered phase $\xi/L$ converges to
zero as the system size increases; in the ordered phase it diverges; at a critical phase point it converges
to finite {\it nonzero} value.

\section{results}
In Ref. \onlinecite{Ray}, the ferromagnetic case of model (\ref{Potts}) is studied. 
It is shown that, if $\lambda>0$, the vortex excitations are suppressed, and the low-temperature phase of the system is QLRO;
as the temperature increases, the system undergoes a BKT transition. If $\lambda<0$, the vortex excitations
are enhanced, which can drives the system to another topological phase: the vortex lattice phase.

\begin{figure}[htpb]
\includegraphics[width=\columnwidth]{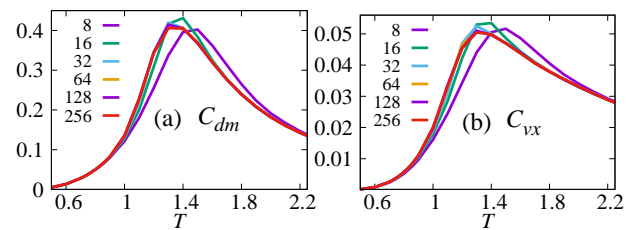}
\caption{(Color online) Specific heat of the generalized antiferromagnetic Potts model (\ref{Potts}), with $\lambda=2$: 
(a), the specific heat of domain wall $C_{dm}$;
(b), the specific heat of vortex $C_{vx}$. }
\label{Cv_pos2}
\end{figure}

\begin{figure}[htpb]
\includegraphics[width=\columnwidth]{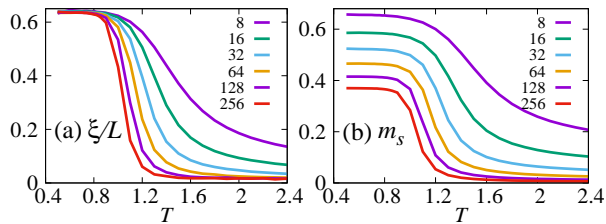}
\caption{(Color online) Correlation length and staggered magnetization of the generalized antiferromagnetic Potts model (\ref{Potts}), 
with $\lambda=2$: (a), the ratio of correlation length to system size, namely $\xi/L$; (b), staggered magnetization $m_s$.}
\label{xims_pos2}
\end{figure}

For the antiferromagnetic case of model (\ref{Potts}), we find that both the two types of topological phases can also be found in it. 
Firstly, for the case of $\lambda>0$, the QLRO is found at the low temperature phase. As shown in Fig. \ref{Cv_pos2}. 
the specific heat $C_{dm}$ and $C_{vx}$ do not show any singularity; 
although they have a peak, the peak is smooth, it does not diverge as the system size increases. 
This is the typical feature of BKT transition. The correlation length $\xi$ and the magnetization $m_s$ also show
the characters of QLRO; the ratio of the correlation length to system size converges to a finite nonzero value in the QLRO phase,
but converges to zero value in the disordered phase, as shown in Fig. \ref{xims_pos2} (a).
 From the figure, we roughly estimate the BKT point to be $T_c=0.75(2)$. 
The magnetization $m_s$ in both the QLRO phase and the disordered phase converges to zero as the system size increases, 
however, the decaying behaviors are different for the two phases, as shown in Fig. \ref{xims_pos2} (b), it decaying algebraically 
in the QLRO phase but exponentially in the disordered phase. At a given temperature below the BKT point, $m_s$ can be fit according to
the following formula
\begin{eqnarray}
m_s \sim L^{y_h-d}, \label{msfit}
\end{eqnarray}
where $y_h$ is the renormalization exponent of the staggered field,
it is related to the critical exponent $\eta$ with $\eta=2(1-y_h)+d$; $d=2$ is the dimension of the system.
By the fitting, we find that in the whole BKT phase, the exponents $y_h$ almost keep invariant, which is $y_h=1.833(1)$;
such value is consistent with the pure antiferromagnetic Potts model, which is critical at the zero temperature, with exponent $\eta=1/3$. 
It means that the vortex term in Hamiltonian (\ref{Potts}) shifts the critical point from the zero temperature to a finite temperature, 
and the zero-temperature `critical point' is enlarged to a `critical phase' (QLRO phase). This result is understandable,
because positive $\lambda$ suppresses the vortex excitations, higher temperature is needed to excite the vortices.
Such BKT transition is not a standard one\cite{XY1,XY2} which has critical exponent $\eta=1/4$ at the BKT point.

\begin{figure}[htpb]
\includegraphics[width=\columnwidth]{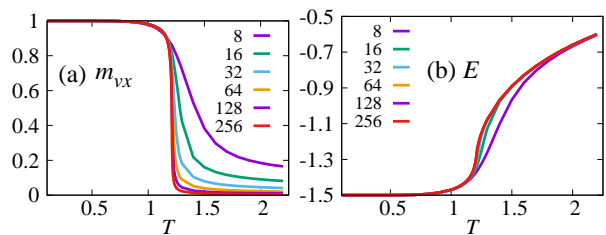}
\caption{(Color online) (a) Vortex magnetization of the generalized antiferromagnetic Potts model (\ref{Potts}), 
with $\lambda=-2$; (b) Internal energy density of the model.} 
\label{mvE-2}
\end{figure}

\begin{figure}[htpb]
\includegraphics[width=\columnwidth]{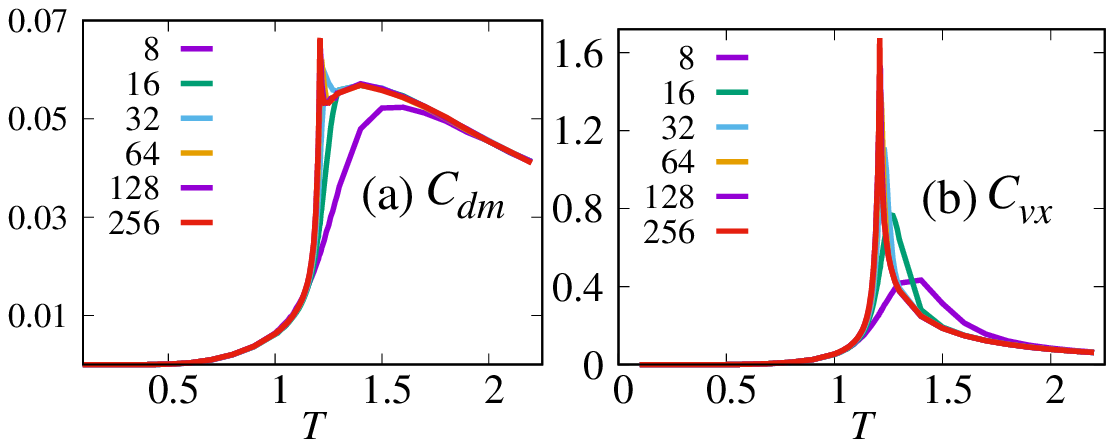}
\caption{(Color online) Specific heat of the generalized antiferromagnetic Potts model (\ref{Potts}), 
with $\lambda=-2$: (a), the specific heat of domain wall $C_{dm}$; (b), the specific heat of vortex $C_{vx}$.}
\label{Cv-2}
\end{figure}

\begin{figure}[htpb]
\includegraphics[width=\columnwidth]{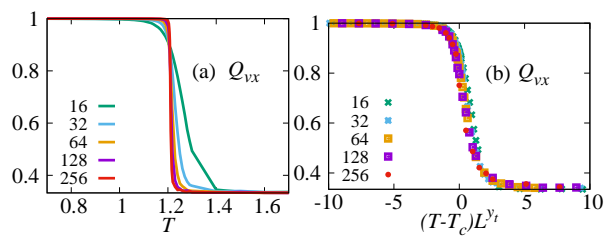}
\caption{(Color online) (a) Binder ratio of the generalized antiferromagnetic Potts model (\ref{Potts}), with $\lambda=-2$; 
(b) Data collapse of the Binder ratio, using $y_t=1$.}
\label{Qv-2}
\end{figure}

Now we turn to the case of negative $\lambda$, in this case the vortex excitations are enhanced, 
thus it is natural to see a vortex lattice phase; a snapshot of such phase is shown in Fig. \ref{PD} (b).
The plot of staggered vortex magnetization $m_{vx}$ is shown in Fig. \ref{mvE-2} (a); 
at the low temperature, $m_{vx}$ is 1, strongly indicates a vortex lattice phase. 
The transition from the vortex lattice phase to the disordered phase is continuous, 
as shown in Fig. \ref{mvE-2} (b), there is no singularity of the curves of the internal energy density.
It should be noted that, such transition is driven by both the vortex energy and the domain wall energy, as shown in Fig. \ref{Cv-2},
both the specific heat $C_{vx}$ and $C_{dm}$ show diverging peaks, although the peak of $C_{dm}$ looks smaller than $C_{vx}$.
The universality class of such transition should be two-dimensional Ising, because it spontaneously breaks the $Z_2$ symmetry.
This can be verified by the behaviors of the Binder ratio. As shown in Fig. \ref{Qv-2} (a), 
The critical point is $T_c=1.210(5)$, obtained by the crossing point of the Binder ratio; the critical value of the Binder ratio, denoted 
by $Q_{vx}^c$, and also obtained by the crossing point, is $Q_{vx}^c\approx0.85$. Such value is universal, it coincides with that of the two dimensional Ising model, which is 0.856(1)\cite{IsingQ}.
Furthermore, we do data collapse for the Binder ratio, using critical exponent $y_t=1$ (which is that of the two dimensional Ising model); 
as shown in Fig. \ref{Qv-2} (b), all the data points collapse to a single line; this also verified that the transition belongs to 
the universality of two dimensional Ising model.

\begin{figure}[htpb]
\includegraphics[width=\columnwidth]{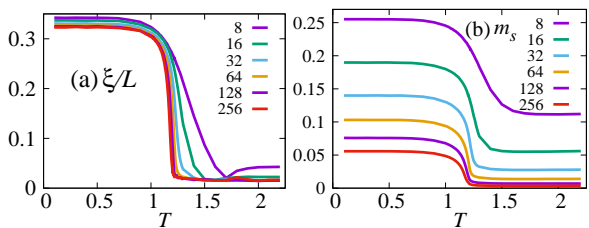}
\caption{(Color online) Correlation length and staggered magnetization of the generalized antiferromagnetic Potts model (\ref{Potts}), 
with $\lambda=-2$: (a), the ratio of correlation length to system size, namely $\xi/L$; (b), staggered magnetization $m_s$.}
\label{xims-2}
\end{figure}
The QLRO phase can also be found when $\lambda<0$.
As shown in Fig. \ref{xims-2} (a), the correlation length in the range $T<=T_c^\prime\approx 1.06(1)$ converges to 
a finite nonzero value, indicating such
phase is QLRO. The behavior of magnetization $m_s$ is similar to the case of $\lambda>0$, as shown in Fig. \ref{xims-2} (b); 
by fitting the data to Eq. (\ref{msfit}), we find the value of critical exponent $y_h$ lies in the range [1.56, 1.50],
accordingly the value of $\eta$ is in the range [0.88, 1.0], this is also very different to the standard BKT transition.
It should be noted that in the temperature range [0, $T_c^\prime$],
the original lattice of the system is in the QLRO phase and the dual lattice is in the vortex lattice phase,
namely the two types of topological phase overlap with each other.

\begin{figure}
 \includegraphics[width=1.0\columnwidth]{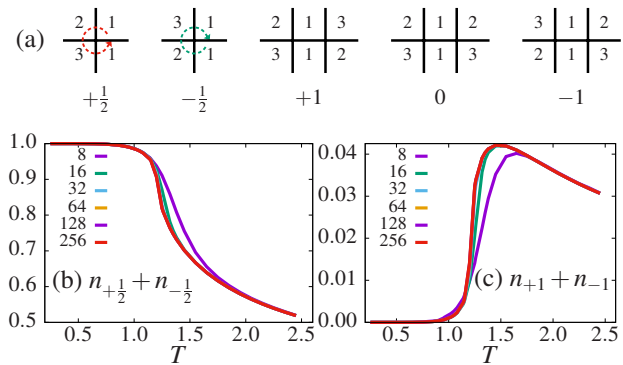}
 \caption{(Color online) (a), vortex excitations in the generalized Potts model; (b) and (c), density of charge half vortices and
  density of charge one vortices of the model with $\lambda=-2$, respectively.}
 \label{vor}
 \end{figure}

The vortex defined in current paper (same to that in Ref. \onlinecite{Ray}) is not exactly the same to
that defined in Refs. \onlinecite{afpvortex1} or \onlinecite{afpvortex2}.
In our opinion, the vortices in current paper can be considered as charge $\pm 1/2$ excitations,
while the vortices defined in Refs. \onlinecite{afpvortex1} or \onlinecite{afpvortex2} are excitations with charge zero or $\pm 1$, 
which can be considered as the combination of the vortices with charge $\pm 1/2$, as shown in Fig. \ref{vor} (a).
In the language of the current, a charge half vortex (defined on the dual lattice) is composed of three pointing in or out arrows 
(current=$\pm 1$) and a line without arrow (current=0), such as Fig. \ref{config} (b). 
In a microstate of the ideal vortex lattice phase,
 all the sites of the dual lattice have such configurations, it can be found that every site is occupied 
by one and only one of the edge with current zero, this is very similar to the full packed dimer model.
 It is known that the full packed dimer model on the square lattice is in a QLRO phase\cite{dimer}, 
thus such mapping gives a theoretical proof that the ideal vortex lattice phase (on the dual lattice) 
can also be a QLRO phase (on the original lattice).
Two neighboring vortices with charge half can compose a vortex with charge zero or charge one; 
once they compose the charge one vortex, it is impossible for them to compose charge one vortex with other neighbors; 
such constraint restricts the state space of configurations with charge one vortices.
Furthermore, it should be noted that the excitation of such type of charge zero or charge one vortex costs the same energy, 
thus the ideal vortex lattice phase is a not only a selection of energy but also the selection of entropy. 

In the case of $\lambda<0$ (and $J<0$), when the temperature is low, the system favors $\pm 1/2$ charges, 
and the charges are in staggered pattern as shown in Fig. \ref{PD} (b), 
there is almost no $\pm 1$ charges which are needed to destroy the QLRO phase.
When the temperature is high enough, the system has enough energy to excite enough $\pm 1$ charges, then it can enter other phase. 
It is interesting that before the system enters the disordered phase, 
it lives in the pure vortex lattice phase (without QLRO) in the temperature range ($T_c^\prime$, $T_c$);
such phase is very subtle, as shown in Fig. \ref{mvE-2} (a), the value of the staggered vortex magnetization $m_{vx}$ in this phase is near but not equal to 1, thus certain amount of $\pm 1/2$ vortices must be combined to $\pm 1$ vortices (see Fig. \ref{vor} (b) and (c)), 
which can destroy the QLRO phase. However, the ratio of the $\pm 1$ vortices should be moderate, 
or else the system will enter the disordered phase. 
In Fig. \ref{vor} (c), it is shown that in the vortex lattice phase the density of charge one vortices is very small,
and the crossing point of the curves is just the critical point. 
Such result means that the vortex lattice phase can also be obtained by suppressing the charge one vortices.
\section{Conclusion}
\label{con}
In conclusion, by means of Monte Carlo simulations with an efficient worm algorithm, 
we studied the generalized antiferromagnetic Potts model (\ref{Potts}). The model can not only exhibit the QLRO phase
and the vortex lattice phase but also the overlap of the two topological phases, which as we know is reported for the first time.
These exotic phases are closely related to the vortex excitations of the model.
Our findings is helpful for understanding the classical topology of physics.
\section{Acknowledgment}
\label{ack}
This work is supported by the National Natural Science Foundation of China under Grant Nos. 11774002 (Ding) and 11625522 (Deng),
and by the Anhui Provincial Natural Science Foundation under Grant No. 1508085QA05 (Ding).


\begin{thebibliography}{widest-label}
\bibitem{XY1}
 V. L. Berezinskii, Sov. Phys. JETP {\bf 32} 493 (1971).
\bibitem{XY2}
 J. M. Kosterlitz and D. J. Thouless, J. Phys. C: Solid State Phys. {\bf 6} 1181-1203 (1973).
\bibitem{clock}
S. K. Baek, P. Minnhagen, and B. J. Kim, Phys. Rev. E {\bf 80}, 060101(R) (2009).
\bibitem{QIsing}
S. V. Isakov and R. Moessner, Phys. Rev. B {\bf 68}, 104409 (2003).
\bibitem{Ising}
S-Z. Lin, Y. Kamiya, G-W. Chern, and C. D. Batista, Phys. Rev. Lett. {\bf 112}, 155702 (2014).
\bibitem{afp3sq}
J. Salas and A. D. Sokal, J. Stat Phys. {\bf 92}, 729 (1998).
\bibitem{afp3layer}
C.-X. Ding, W.-A. Guo, Y. Deng, Phys. Rev. B {\bf 90}, 134420 (2014).
\bibitem{Nijs}
M. P. M. den Nijs, M. P. Nightingale, and M. Schick, Phys. Rev. B {\bf 26} 2490 (1982).
\bibitem{afpvortex1}
J. Kolafa, J. Phys. A: Math. Gen. {\bf 17}, L777 (1984).
\bibitem{afpvortex2}
C. Moore, M. G. Nordahl, N. Minar, and C. R. Shalizi, Phys. Rev. E {\bf 60}, 5344 (1999).
\bibitem{Ray}
S. Bhattacharya and P. Ray, Phys. Rev. Lett. {\bf 116}, 097206 (2016).
\bibitem{SW}
R. H. Swendsen and J. S. Wang, Phys. Rev. Lett. {\bf 58}, 86 (1987).
\bibitem{Wolff}
U. Wolff, Phys. Rev. Lett. {\bf 62}, 361 (1989).
\bibitem{WSK}
J. S. Wang, R. H. Swendsen, and R. Koteck\'{y}, Phys. Rev. Lett. {\bf 63}, 109 (1989).
\bibitem{Geo}
C. Dress and W. Krauth, J. Phys. A {\bf 28}, L597 (1995); J. R. Heringa and H. W. J. Bl\"{o}te, Physica A {\bf 232}, 369 (1996); 
J. R. Heringa and H. W. J. Bl\"{o}te, Phys. Rev. E {\bf 57}, 4976 (1998).
\bibitem{IsingQ}
G. Kamieniarz and H. W. J. Bl\"{o}te, J. Phys. A: Math. Gen. {\bf 26}, 201 (1993).
\bibitem{dimer}
M. E. Fisher and J. Stephenson, Phys. Rev. {\bf 132}, 1411 (1963).
\end{thebibliography}
\end{document}